# Solar Activity and Transformer Failures in the Greek National Electric Grid

*ZOIS Ioannis Panayiotis*

Testing Research and Standards Centre, Public Power Corporation
9, Leontariou Street, GR 153 51, Kantza, Pallini, Athens, Attica,
GREECE

Email: i.zois@dei.com.gr

# Abstract

**Aims**
We study both the short term and long term effects of solar activity on the large transformers (150kV and 400kV) of the Greek national electric grid.

**Methods**
We use data analysis and various statistical methods and models.

**Results**
Contrary to common belief in PPC Greece, we see that there are considerable both short term (immediate) and long term effects of solar activity onto large transformers in a mid-latitude country like Greece. Our results can be summarized as follows:
1. For the short term effects: During 1989-2010 there were 43 "stormy days" (namely days with for example Ap ≥ 100) and we had 19 failures occurring during a stormy day plus or minus 3 days and 51 failures occurring during a stormy day plus or minus 7 days. All these failures can be directly related to Geomagnetically Induced Currents (GIC's). Explicit cases are presented.
2. For the long term effects, again for the same period 1989-2010, we have two main results:
   (i) The annual number of transformer failures seems to follow the solar activity pattern. Yet the maximum number of transformer failures occur about half a solar cycle after the maximum of solar activity.
   (ii) There is statistical correlation between solar activity expressed using various newly defined long term solar activity indices and the annual number of transformer failures. These new long term solar activity indices were defined using both local (from geomagnetic stations in Greece) and global (planetary averages) geomagnetic data. Applying both linear and non-linear statistical regression we compute the regression equations and the corresponding coefficients of determination.

# 1. Introductory Theory - Motivation

Solar Coronal Mass Ejections (CME's) and/or high speed streams (Co-rotating Interaction Regions or CIR's) of the solar wind originating from a region of weak magnetic field on the Sun's surface cause disturbances in the interplanetary medium. These disturbances in the interplanetary medium in turn cause temporary disturbances of the earth's magnetosphere called Geomagnetic Storms. A geomagnetic storm is caused by a solar wind shock wave and/or cloud of magnetic field which interacts with the Earth's magnetic field. The increase in the solar wind pressure initially compresses the magnetosphere and the solar wind's magnetic field will interact with the Earth's magnetic field and transfer an increased amount of energy into the magnetosphere. Both interactions cause an increase in movement of plasma through the magnetosphere (driven by increased electric fields inside the magnetosphere) and an increase in electric current in the magnetosphere and ionosphere. During the main phase of a geomagnetic storm, electric current in the magnetosphere create magnetic force which pushes out the boundary between the magnetosphere and the solar wind.

The frequency of geomagnetic storms increases and decreases with the sunspot cycle. CME driven storms are more common during the maximum of the solar cycle and CIR driven storms are more common during the minimum of the solar cycle. In 1930, Sydney Chapman and Vincenzo C. A. Ferraro wrote an article, *A New Theory of Magnetic Storms*, (see Chapman et all 1930) that sought to explain the phenomenon of geomagnetic storms. They argued that whenever the Sun emits a solar flare it will also emit a *plasma cloud*. This plasma will travel at a velocity such that it reaches Earth within 113 days. The cloud will then compress the Earth's magnetic field and thus increase this magnetic field at the Earth's surface.

There are several space weather phenomena which tend to be associated with a geomagnetic storm or are caused by a geomagnetic storm. These include: Solar Energetic Particle (SEP) events, geomagnetically induced currents (GIC), ionospheric disturbances which cause radio and radar scintillation, disruption of navigation by magnetic compass and auroral displays at much lower latitudes than normal. In this article we shall focus on GIC's and their effects on electric grids, in particular large transformers. GIC's are the ground level manifestations of space weather (namely geomagnetic storms) and they are driven by the variations of the terrestrial magnetic field (temporary disturbances of the earth's magnetosphere) via Faraday's Law of Induction.

The greatest GIC problems occur at high latitudes in or near the auroral zones. In these areas the geomagnetic storms are most intense and frequent since the ionospheric source is typically a localised electrojet. Countries at mid-latitude, such as Greece (latitude approximately $35^o$ N – $41^o$ N), are located far from the magnetic pole and do not experience the same severity of geomagnetic disturbances as Canada or Scandinavia. GICs and their potential threat to network integrity are relatively unknown at mid latitudes. Until recently it was believed that electricity networks in mid-latitude regions are not affected by GICs. Our research addresses this assumption, and provides insight into the incidence of GICs and the performance of the Greek electricity transmission network.

The effects of GIC's on transformers are briefly the following: GIC's superimpose a direct current onto the AC waveform. Yet although this direct current is usually rather small, it can shift the (B-H) curve outside the linear design operation region resulting in highly distorted output waveform, very high magnetic fluxes and half-cycle saturation (see for example Bossi et all 1983, Takasu et all 1994, Minhas et all 1999, Price 2002). The magnetic flux leakage from the core of the transformer may lead to overheating, gassing, shutdown and even catastrophic failure. Moreover the distorted output wave form can create higher harmonics, can cause the triggering of the protective relays (Buchholz etc) and various control assets may be switched out. In addition to all the above the magnetizing current can get outside the design specifications leading to increased VAr consumption and voltage instability. Let's describe the situation with slightly more details:

GIC's are driven by the variation of the (terrestrial) magnetic field, so they are characterised by slow changes. A transformer core saturates under GIC bias, causing it to operate in the extremely non-linear portion of the core steel magnetisation (B-H) curve. This quasi-direct current (GIC) flowing through a transformer off-sets the power frequency magnetisation curve so that the magnetic circuit operates asymmetrically on the (B-H) curve. Transformers are designed to operate in the linear portion of the curve with only a small magnetising current, without approaching the non-linear regions of core saturation. Even small GIC's can produce sufficient off-set for the transformer to reach half-cycle saturation of the ac current that is in the same direction as the GIC's. As the core saturates the permeability tends to 1 and the flux fills the whole space of the winding, so it is very different from the flux distribution under normal conditions. Normal leakage flux is controlled by non-magnetic (stainless steel or aluminium) shields, shunts or cheek plates, the dimensions of which are critical for controlling eddy current losses and heating. Under normal conditions, with most flux in the magnetic core, the non-magnetic plates carry little flux and generate only small losses. Under half-wave core saturation, the losses in some parts of the leakage flux shields may increase significantly, causing localised heating.

Half-wave saturation also increases the magnetising current and distorts the magnetisation current waveform (see for example Takasu et all 1994) . The distorted magnetising current has harmonics that correlate well with the GIC. Without adequate control of the flux under saturation conditions, local heating in parts of the transformer may not be cooled effectively, leading to rapid temperature increase. The intensity of overheating depends on the saturation flux paths, cooling flow and the thermal condition or loading of the transformer. Overheating causes the breakdown of oil and paper insulation, leading to gassing that can be detected and analysed by dissolved gas analysis (DGA).

When a large number of transformers experience saturation due to GICs, the system's reactive power demand increases significantly compared with the total load supplied by the transformer. Reactive power demands of this magnitude can cause severe system voltage excursions (see Schrijver, Mitchell 2013). At the same time, the change in size of the ferromagnetic material (magnetostriction) between saturated and unsaturated states at 100 times a second (at 50 Hz) causes heating, noise and mechanical vibration damage, thus possibly leading to many "mechanical" failures (for example broken switch axes) on top of the "electric" failures.

In the scientific literature (see for example Coetzee & Gaunt 2007; Gaunt & Kohen 2009) there have been reports that GIC's can cause many types of failures on transformers like: Insulation failure, tank shunting effects, local heating, rapid temperature increase causing the breakdown of the oil and paper insulation, thus leading to gassing (which can be analysed by dissolved gas analysis). The DGA records of transformers in S. Africa for example indicate that deterioration continues after the initial damage caused by GIC's, affected by transformer loading and possibly other stresses (see Kohen et all 2002). *A transformer might fail only months after the initial damage* and, unless frequent DGA samples have been taken, could appear to fail for "unknown causes" or a subsequent stress incident. Failure of transformers is often ascribed to damage of internal insulation by external over-voltages or network faults. Other significant causes of transformer failures are settling in and aging. Copper sulphide formation in transformers or reactors with corrosive oils has recently become a concern. The coincident onset of gas generation in widely separated transformers of different ages, the relative "exposure" of the transformers to GICs, and the similarities between the nature of damage and its timing all indicate that the system-wide effect of GICs is likely to be a more significant initiator of the failures than overloading, over-voltages, system faults, copper sulphide formation or manufacturing defects. Some of these other factors may contribute to failure by weakening transformers and explain why some failed and others did not. The damage in all the transformers inspected appears to be initiated by local overheating. The DGA results are consistent with low temperature degradation of insulation. The levels of dissolved gases are below levels generally considered to be significant, and this is consistent with localised overheating.

The nature of faults caused by low temperature local overheating in windings is such that much of the evidence at the site of a fault would be obscured by subsequent disruptive failure. It is unlikely, therefore, that inspections after a disruptive fault will identify evidence of low temperature local overheating at the fault position. Unless overheating caused by leakage flux established by GICs is specifically considered, which is unlikely based on current knowledge of the problem, GICs will not be reported as a cause of faults. Does damage caused by low temperature local overheating always leads to failure and, if so, how quickly? Once core saturation by the GIC is removed, thermally damaged paper insulation will be less robust than before the event. Another mechanism of further damage must be considered, such as by leakage currents, partial discharge or reduced heat transfer through the damaged paper, causing further local heating to an extent that degradation continues. Such mechanisms could explain why gas generation and DGA levels decrease when the transformer loading is reduced. Eventually, depending on the extent of the initial damage, the presence of air and water in the transformer, and operating stresses, the damaged insulation in a small area will fail, even if the DGA levels are significantly below those usually indicating incipient failure. However, since the initial damage and the subsequent operating conditions are variable, the breakdown level or likelihood of failure is difficult to determine. *In the cases reported in S. Africa and elsewhere, the DGA indicated the onset or increase of damage coincident with a severe storm. Thereafter, some transformers failed quite quickly (after a few days from the incident—this is one of the reasons for the choice of 3 and 7 days time scales after a storm for immediate failures in section 5-- others took months, and some might survive until gradual deterioration, a severe system fault, overvoltage or* *another GIC event causes the damage to be extended to the point of failure.* Accordingly, any transformer exhibiting an increase in levels of dissolved gasses that indicate paper degradation after a geomagnetic event should be considered as distressed. The nature or location of GIC initiated damage is not the same for all the transformers. Designs differ and transformers made to the same design are not always manufactured identically. Similar GICs through similar transformers might cause slightly different patterns of leakage flux, and the weakest part of one transformer's insulation might differ from another's. The most likely failure point will be where leakage flux creates a condition that exceeds the local cooling capacity. In extreme cases, part of the tank or core might melt, but in most cases damage might only be observed in a winding or a lead, as in the cases reported here. *The GIC's should not be thought of as a single cause of a fault, but as a stress that exposes relative weaknesses, which become localised hot spots and eventually lead to failure. Thus in many cases these failures can be attributed to "aging" or "manufacturing defects".*

*GIC's have cumulative and long term effects on transformers.* Extreme GIC events do not occur only at the peak of or late in a solar cycle and final failure may not easily be linked to the solar cycle or a specific storm. Yet based on these responses to GICs, it would be expected that transformer failures should increase during periods following geomagnetic storms. Such a trend is evident for example from an analysis of failures of large transformers, all over 230 kV, in S. Africa and N. America. The same holds true in our case (Greece) where we observe an important increase in the number of failures 7 days after a storm.

There are some interesting implications for the analysis and reporting of failures. Excluding GICs from consideration results in all faults being ascribed to other causes. Subsequent analysis of fault data would show that GICs are not reported as a cause of faults, further obscuring the possible underlying processes of damage initiation. One of the main motivations for this research is that utility engineers (PPC or IPTO in our case here concerning Greece) should start recognising that a number of transformer failures are due to GICs, based on evidence available from storm occurrences, network analysis, incident reports and damage inspections. This in turn can attach liability to transformer suppliers with less access to the relevant data. A manufacturer that could show that failures might be associated with GICs, but that GICs were not considered or identified in specifications or incident reports by the utility engineers, could avoid or reduce any liability for damage. Avoiding damage by GICs requires that transformers be designed and built so that leakage flux resulting from saturation will not cause local overheating.

The second motivation (and perhaps more important) for this work comes from mitigation: Mitigation can be implemented by suitable flux shunts, adequate clearances between the tank and core, aligning conductors so that leakage flux does not produce significant eddy currents, improving the cooling where local heating might be expected, and other techniques already used by manufacturers. Transformers have been designed, built and tested to survive saturation by very high GICs. Some utilities already specify that transformers must withstand the effects of large GICs, with compliance demonstrated by finite element modelling of the flux produced. Design reviews for transformers in less exposed situations could similarly take into account the effects of GICs at levels appropriate to the network. Temporary network reconfiguration can be used to reduce the magnitudes of GICs, and might be a suitable way to reduce transformer stress in networks that are only infrequently exposed to severe geomagnetic storms, such as outside the auroral zones. Where many transformers are already installed without specified GIC capability, temporary network reconfiguration during storms may be the most effective short-term mitigation procedure. Factors to be considered include the reliability of the reconfigured network, the costs of modified power despatch, the transformer risk associated with the redistributed GICs, and the conditions under which reconfiguration will be implemented. Long-term mitigation can be achieved by installing series of capacitors in long transmission lines to block the quasi-DC GICs, with possible benefits for power flow capacity or even with GIC-blocking equipment.

## 2. Transformer Failures in Greece – The collection of the relevant data

The Greek national electric grid contains a few dozens of mega transformers (400kV) and a few hundreds of large transformers (150kV) plus a number of auxiliary transformers (starting from 15 kV).

The Independent Power Transmission Operator (IPTO or ADMIE in Greek) is the Greek independent power transmission operator, a fully subsidiary company of PPC (the Greek Public Power Corporation). IPTO provided data for mega and large transformer failures (namely failures for 400 kV and 150 kV transformers) throughout Greece from 1989-2010.

According to the IPTO (and PPC) policy *a recorded transformer failure* is any transformer malfunction having as a result the transformer shut down; the duration of the shut down can be anything ranging from 1 min up to complete destruction and total replacement of the transformer.

Again according to IPTO (and PPC) policy, recorded transformer failures are divided into 2 categories: These caused by human errors and all the rest. The former are recorded separately and they are used for personnel training and personnel evaluation. IPTO provided data only for the later (with the remark that failures due to human error constitute a negligible amount). Hence we shall only consider, to begin with, *recorded transformer failures caused by any reason except human error.*

Going through the available data, there are some transformer failures which are obviously irrelevant to GIC's. These are failures due to:
- earthquakes,
- falling of trees or other objects on the transformer (for example due to strong winds),
- short circuits due to animals (for example in Greece it is relatively common to have short circuits caused by ferrets or even occasionally birds),
- lightnings,
- extreme weather conditions (heat or frost).

These failures have been excluded from our study. All the remaining failures are included, no matter if they are characterized as "mechanical" or "electrical" or whatever else. (For example failures due to "aging" or "manufacturing defects" are included since there are reports on their relevance--see the previous paragraph). *These failures in the sequel will be called simply transformer failures (TF's). We would like to mention though as an important note that these obviously irrelevant to GICs transformer failures are about 10% of the total amount of recorded failures hence even if we had included them the statistical results would not have been much different. TF were grouped on an annual basis.*

Let us now explain and justify our strategy and methods. We start by exhibiting the problems and limitations: Looking at the available data for TF, it is very hard indeed to state whether a specific failure is explicitly and exclusively due to GIC (taking in mind the previous paragraph where we explained that GIC's aggravate transformers and can cause cumulative effects). In addition to this, there are no devices measuring GIC's or any other sort of relevant monitoring equipment anywhere in Greece. And finally, repairing technicians—who are the people who complete the failure reports—admittedly have no training or experience whatsoever on GICs and failures related to them.

We tried to overcome these limitations with the following approach: First we divided the effects of solar activity onto transformers in 2 categories: Short term (or immediate) effects and long term effects (for the definition of the immediate or short term effects see the next paragraph).

For the immediate effects things were a bit complicated. Our task was to try to select possible immediate effects of solar activity on transformers by looking at TF data and reports written by technicians which are totally unfamiliar with GIC's and the problems they create. Given our restrictions described above, what we did then was this: For the period 1989-2010 for which we have available data, we picked up the "stormy days", namely days with say $Ap \geq 100$ (this roughly corresponds to G2 and higher in the NOAA G-scale). For these stormy days then, we counted on how many of them we had a TF. In addition we counted how many failures we had during a stormy day or 3 days prior or 3 days after a stormy day; this is in order to take into account effects like *sudden commencements of storms where rapid variations of the terrestrial magnetic field occur*. We did the same for a 7 day period (before or after a stormy day). The rationale behind this is that TF's occurring during a stormy day (or "around" a stormy day) are very possible to be caused by GIC's indeed. Thus we give the relevant frequency (frequency of occurrence of failures during a stormy day) along with explicit examples which we briefly present and analyse. This provides an answer the question of the assessment of the immediate or short term effects of solar activity onto transformers.

The later, namely long term effects of solar activity onto transformer failures is-- in principle at least--perhaps simpler to deal with and quantify: We use various analytic and statistical methods, in particular linear and non-linear statistical regression and correlation. As an independent variable we use some newly defined long term solar activity index and as a dependent variable we use the annual number of transformer failures. One basic problem we had to overcome here at the very start was to create a suitable long-term solar activity index since there was no one readily available.

## 3. Solar activity (long term indices) data

In order to assess-study the long term effects of solar activity on transformers, the idea is to try to statistically correlate transformer failures with solar activity. The subtlety here is that we need some "indices" with the same time scale for these two groups of data. From the transformer failures' side, things are pretty straightforward, the annual number of transformer failures throughout Greece is probably the most reasonable index (counting the number of transformer failures on a monthly basis may have some logic yet it is not very convenient; any shorter time scale than that, weekly, daily or hourly basis, is clearly senseless). On the solar activity side things are not so straightforward. A very reasonable index measuring solar activity is the number of sunspots, given on an annual basis. The number of sunspots follows well (in fact partly defines) the 11-year solar activity cycle. The problem with this index (annual number of sunspots) though is that it is "indirect" for our purposes. We are primarily interested in *earth affecting solar activity* (and not just solar activity in general) and for this purpose the number of sunspots is relevant but not directly relevant; for example in sunspot regions flares or CME's tend to be created but whether they affect the earth and up to what extend, depends on the size of flare or CME, the relative position (direction and magnitude) of the flare with the current position of the earth. Thus a more direct index should be directly related to the variations of the magnetic field of the earth caused by the ejected plasma from the sun. After all the creation of GIC's (which is what affects transformers) is due to the Faraday Law in classical electromagnetism which relates electric fields and currents with variations of the magnetic field. Now there are indeed certain indices measuring these variations of the terrestrial magnetic field, the most important and well-known ones are the Ap index (and its logarithmic version the Kp index along with some variations) and the Dst index (see Adams 1892, Akasofu & Chapman 1964, Akasofu & Chapman 1972, Bartels et al. 1939, Broun 1861, Cahill 1966, Campbell 2002, Carlowicz & Lopez 2002, Chapman 1935, Chapman 1952, Crooker & Siscoe 1981, Frank 1970, Fukushima & Kamide 1973, Kertz 1958, Kertz 1964, Rangarajan 1989, Sugiura 1991, Sugiura & Hendrics 1967). The problem with these indices however is that they are short term indices, they are daily planetary averages from about a dozen stations on the globe. We need a long term (on an annual basis) index of the earth affecting solar activity. So the picture is this: On the one hand the annual number of sunspots is a good long term index of solar activity but it is characterized as indirect for our purpose whereas on the other hand the Ap and Dst indices are direct but short term indices (see Hamilton 1986, Haymes 1971, Hess 1968, Kivelson & Russell 1995, Langel et al. 1980, Lehtinen & Pirjola 1985, Mayaud 1980, Moos 1910, Parks 1991, Rostoker 1972, Shelley 1979, Smith et al. 1981, Van Allen 1983, Vestine et al. 1947, Walt 1994, Williams 1981). It is not easy to get a direct index out of the number of sunspots but there are a number of possibilities to define some long term (on an annual basis) solar activity indices using the Ap and Dst data: One can count the number of "disturbed" days per year, namely annual number of days with say Ap $\geq$ 40 and Dst $\leq$ -40. The value 40 was chosen from the NOAA Boulder Colorado station scale table and the NOAA Space Weather Scale for Geomagnetic Storms (NOAA G-scale): it seems a reasonable choice since the A-index is the daily average of the equivalent a indices and a value of a = 40 corresponds to roughly a situation *between G0 and G1* (G0 means no practical effects, G1 means minor practical effects on man made systems on earth). Table 1 indicates the correspondence between the K-index, the a-index, the variations of the terrestrial magnetic field ΔB and the NOAA G-scale:

Table 1: Boulder Colorado station scale table

| K-index | a-index | Boulder Colorado USA observatory ΔB measurements (horizontal magnetic field variations in **nT**) | NOAA G-scale |
|---|---|---|---|
| 0 | 0 | 0-5 | G0 |
| 1 | 3 | 5-10 | G0 |
| 2 | 7 | 10-20 | G0 |
| 3 | 15 | 20-40 | G0 |
| 4 | 27 | 40-70 | G0 |
| 5 | 48 | 70-120 | G1 |
| 6 | 80 | 120-200 | G2 |
| 7 | 140 | 200-330 | G3 |
| 8 | 240 | 330-500 | G4 |
| 9 | 400 | > 500 | G5 |

This long term index has the disadvantage that we count the disturbed days but not the severity of the solar storms. Another option would be to count the annual sum of the Ap and Dst indices, in this case we take into account both the number of storms and their intensity. A third option is to consider the annual average of the Ap and Dst indices. The difference between the annual sum and the annual average is the division by 365 (or 366) and for linear models this makes no difference (as one can easily check by looking at the relevant formulae) but it does make some (little as we shall see in one example) difference in non-linear models. Our final option was to consider the annual mean of the daily sum of the 3 h K-index of the local Penteli Magnetic Observatory near Athens. This last option makes use of local geomagnetic data (and not planetary averages like the Ap and Dst).

If we restrict our study to linear regression and correlation (thus ignoring non-linear phenomena, which is not too bad as a first approximation), it follows easily from the relevant formulae (mathematical statistics) that instead of studying for example the linear correlation between two variables say X and Y, one can equivalently study their scalar multiples aX and bY, (where a, b real scalars), the results remain the same (the correlation coefficient does not change). In this case if we simply divide the annual sum of Ap and Dst by the annual number of days (365), we get the annual average of Ap and Dst. Hence for linear regression and correlation, the annual sum of Ap (and Dst) can be replaced by the annual average of these indices.

In Table 2 we give the solar activity long term indices data: SSN stands for (Annual) Sunspot Number; SSN data were provided by NASA

http://solarscience.msfc.nasa.gov/greenwch/spot_num.txt

All the rest of the data were provided by the UK Solar System Data Centre UKSSDC. Dst data are only available until 2008. For 2009 and 2010 we used provisional data.

Table 2: Long Term Solar Activity Data

| No | Year | SSN | Annual Ap mean | Annual Dst mean (absolute value) | Annual No of days with Ap ≥ 40 | Annual No of days with Dst ≤ -40 |
|---|---|---|---|---|---|---|
| 1 | 1989 | 1893 | 19 | 30 | 38 | 106 |
| 2 | 1990 | 1707 | 16 | 21 | 23 | 68 |
| 3 | 1991 | 1749 | 23 | 31 | 54 | 112 |
| 4 | 1992 | 1134 | 16 | 26 | 23 | 76 |
| 5 | 1993 | 657 | 15 | 19 | 19 | 38 |
| 6 | 1994 | 358 | 18 | 23 | 39 | 65 |
| 7 | 1995 | 210 | 13 | 17 | 10 | 35 |
| 8 | 1996 | 103 | 9 | 11 | 0 | 2 |
| 9 | 1997 | 258 | 8 | 14 | 5 | 20 |
| 10 | 1998 | 770 | 12 | 17 | 16 | 38 |
| 11 | 1999 | 1118 | 13 | 13 | 12 | 35 |
| 12 | 2000 | 1434 | 15 | 19 | 23 | 43 |
| 13 | 2001 | 1331 | 13 | 18 | 19 | 43 |
| 14 | 2002 | 1249 | 13 | 21 | 17 | 64 |
| 15 | 2003 | 763 | 22 | 22 | 35 | 43 |
| 16 | 2004 | 485 | 13 | 12 | 14 | 22 |
| 17 | 2005 | 357 | 13 | 16 | 21 | 40 |
| 18 | 2006 | 182 | 8 | 12 | 4 | 14 |
| 19 | 2007 | 90 | 7 | 8 | 0 | 0 |
| 20 | 2008 | 34 | 7 | 8 | 0 | 2 |
| 21 | 2009 | 37 | 4 | 3 (provisional) | 0 | 1 (provisional) |
| 22 | 2010 | 198 | 6 | 6 (provisional) | 3 | 4 (provisional) |

We give below the relevant graph: Fig 1 gives the graphs of all these long-term indices (which are scaled in order to match the SSN values). We observe that all these indices have the general shape of the SSN curve, they exhibit roughly the same periodicity as SSN (11 year solar cycle).

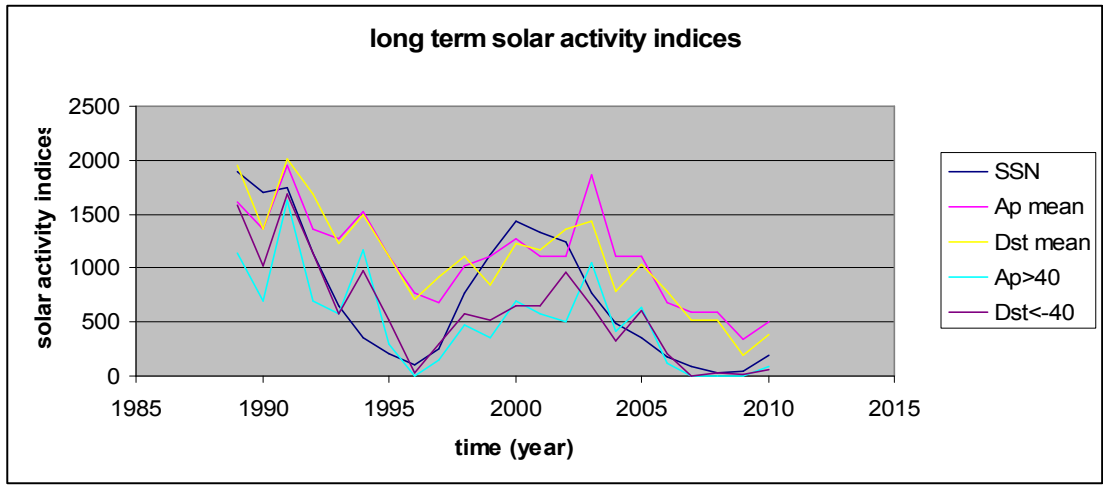

Fig 1: Long Term Solar Activity Indices (all scaled except SSN) vs Time graph

## 4. Long term effects

### I. Global indices

For the long term effects we shall use various statistical methods and tests. The simplest thing to do in order to start our study would be to run a linear regression model (see Zois 2011) between transformer failure number (TFN) and sunspot number (SSN). The sunspot number (SSN) will be our independent variable X whereas the transformer failure number (TFN) will be our dependent variable Y. The *(linear)* regression equation we find is

$$Y = -0.01\, X + 48.09.$$

The slope is -0.01, the y-intercept equals 48.09, the corresponding correlation coefficient is r = -0.52 and the coefficient of determination equals $r^2 = 0.27$. These parameters were estimated using the ordinary least squares method (with the use of R and Microsoft Excel). We give below in Fig 2 the relevant scatter plot along with the linear regression line:

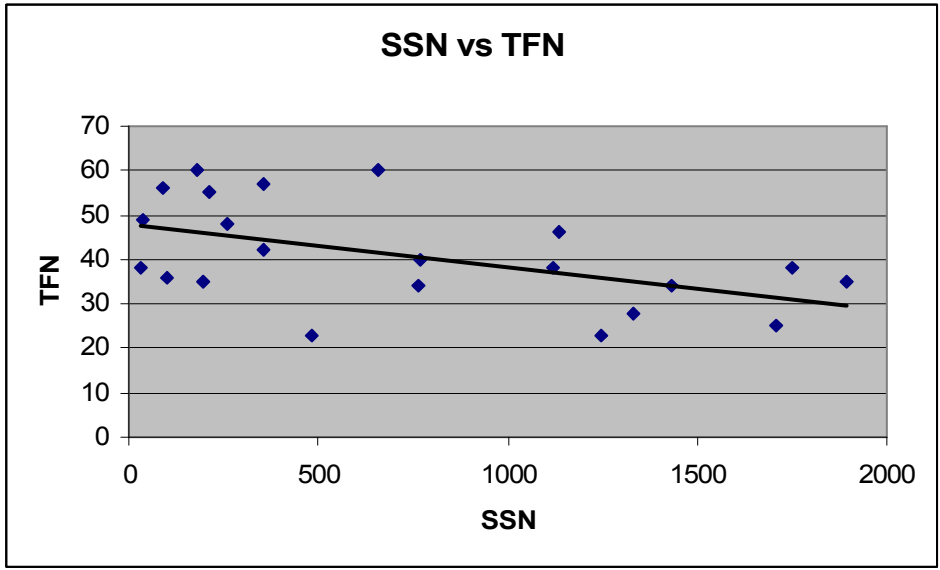

Figure 2: SSN vs TFN scatter plot

One interesting feature here is that the slope is negative (negative correlation). The explanation of this fact is the following: In Fig 3 below we plot the SSN and TFN curves vs time in a single graph and we observe that (although the TFN has considerable noise and fluctuates), there is a "phase delay" between the two curves which is roughly equal to 4-5 years, namely approximately half a solar cycle. For a clearer visual exhibition of this phase delay we have also plotted the smoothed (and scaled) TFN curve. This means that the maximum number of failures occur towards the end of the solar cycle when solar activity attains its minimum. This is due to the cumulative nature of solar activity effects on transformers in the long term: There is a building up phenomenon, as time passes solar activity aggravates transformers. (In Fig 3 below we have scaled the TFN curve for clearer visual exhibition). So Fig 3 verifies the negative sign of the correlation coefficient of the linear approximation.

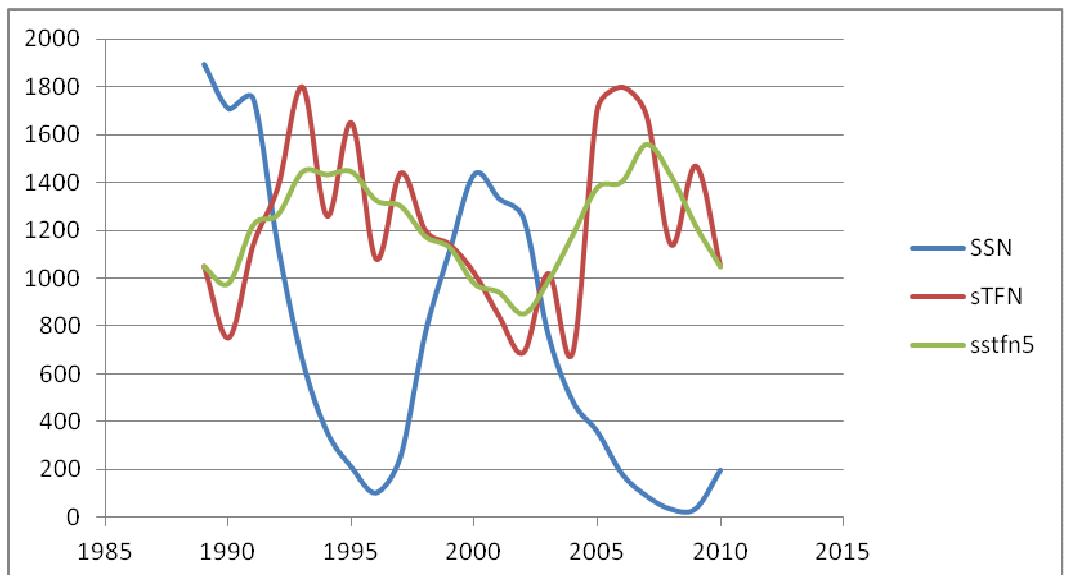

Figure3: (scaled) TFN and SSN vs Time graph. The blue curve is SSN, the red curve is the (scaled) TFN and the green curve is the smoothed (and scaled) TFN.

We can perform a hypothesis testing (Daniel & Terrell 1995) using statistical inference to see if there is (linear) correlation between our variables X (SSN) and Y (TFN). We want to test the null hypothesis $H_0: \rho = 0$ (no correlation, $\rho$ denotes the population correlation coefficient, r which denotes the sample correlation coefficient is an estimator of $\rho$) against the alternative hypothesis $H_1: \rho \neq 0$ (there is correlation). We choose the a = 0.05 confidence level and we use Student's t-variable

$$t = r\sqrt{\frac{N-2}{1-r^2}}.$$

This is a 1-sided test, hence we look at $t_{0.975}$ with N-2 = 22-2 = 20 degrees of freedom. The critical value then of t is 2.086. Since by the above formula we compute t = -2.724 (namely in absolute value larger than the critical value), we reject the null hypothesis. Thus there is correlation.

The next order of business is to try some non-linear models (with respect to the variable X). We have studied the following functional models: logarithmic, polynomial (degrees 1 - 6), hyperbolic and exponential (Barlow 1993, Bjorck 1996, Cohen et al. 2003, Draper & Smith 1998, Goodall 1993, Nievergelt 1994, Pedhazur 1982, Pindyck and Rubinfeld 1998, Kaw and Kalu 2008, Venables & Smith 2012):

*Logarithmic:*

$$Y = a \ln X + b$$

where a, b real parameters.

*Polynomial* of degree 2 up to 6 in X:

$$Y = b + a_1 X + a_2 X^2 + a_3 X^3 + a_4 X^4 + a_5 X^5 + a_6 X^6$$

where $a_1, a_2, a_3, a_4, a_5, a_6$ and b real parameters.

*Hyperbolic:*

$$Y = a X^b$$

where a, b real parameters.

*Exponential:*

$$Y = a e^{bX}$$

where a, b real parameters.

We estimate the parameters using again ordinary least squares. We apply all these functional models to all our long term indices (see Zois 2012). The results are summarized in the following table which contains the equations and the corresponding coefficient of determination $R^2$:

Table 3: Summary of Regression results (global geomagnetic data)

| Index | Function | Equation | $R^2$ |
|---|---|---|---|
| SSN | linear | $Y = -0.01 X + 48.09$ | 0.27 |
| | ln | $Y = -4.12 \ln X + 65.92$ | 0.19 |
| | poly | $Y = 36.872 + 0.2032X - 0.0011X^2 + 2E{-}06X^3 - 2E{-}09X^4 - 9E{-}13X^5 - 2E{-}16X^6$ | 0.35 |
| | hyp | $Y = 75.09 X^{-0.11}$ | 0.20 |
| | exp | $Y = 47.01 e^{-0.0002 X}$ | 0.26 |
| Ap mean | linear | $Y = -0.6416 X + 49.16$ | 0.078 |
| | ln | $Y = -7.49 \ln X + 59.39$ | 0.082 |
| | poly | $Y = 353.07 - 195.63X + 47.664X^2 - 5.7355X^3 + 0.3622X^4 - 0.0115X^5 + 0.0001X^6$ | 0.123 |
| | hyp | $Y = 62.54 X^{-0.19}$ | 0.081 |
| | exp | $Y = 47.84 e^{-0.0153 X}$ | 0.069 |
| Dst mean | linear | $Y = -0.3375 X + 46.539$ | 0.05 |
| | ln | $Y = -4.2932 \ln X + 52.471$ | 0.04 |
| | poly | $Y = 39.998 + 10.631X - 3.9763X^2 + 0.5504X^3 - 0.0348X^4 + 0.001X^5 - 0.00001X^6$ | 0.23 |
| | hyp | $Y = 52.831 X^{-0.1099}$ | 0.04 |
| | exp | $Y = 45.888 e^{-0.089 X}$ | 0.04 |
| Annual no of days with Ap $\geq$ 40 | linear | $Y = -0.2007 X + 44.33$ | 0.06 |
| | ln | --- | |
| | poly | $Y = 43.942 + 5.0937X - 1.2764X^2 + 0.1086X^3 - 0.0042X^4 + 8E{-}05X^5 - 5E{-}07X^6$ | 0.14 |
| | hyp | --- | |
| | exp | $Y = 42.451 e^{-0.046 X}$ | 0.05 |
| Annual no of days with Dst $\leq$ -40 | linear | $Y = -0.1101 X + 44.87$ | 0.08 |
| | ln | --- | |
| | poly | $Y = 50.453 - 4.9074X + 0.5092X^2 - 0.0192X^3 + 0.0003X^4 - 2E{-}06X^5 + 7E{-}09X^6$ | 0.22 |
| | hyp | --- | |
| | exp | $Y = 43.215 e^{-0.024 X}$ | 0.07 |

In the non-linear case, when one uses a scalar multiple of the variable this can make some difference, hence it makes sense to try to use the annual sum of the Ap or Dst index instead of the annual mean. However as we shall see in our case the difference is small, practically non-existing. As an example we consider the annual Dst sum (absolute value, instead of the annual Dst mean):

**Annual Dst sum (absolute value) vs TFN**

*Linear:*
$$Y = -0.3309 X + 46.551$$
with $R^2 = 0.05$

*Logarithmic:*
$$Y = -4.3059 \ln X + 57.912$$
with $R^2 = 0.04$

*Polynomial:* Among polynomials in X with degrees from 2 up to 6, the largest coefficient of determination was given using a polynomial of degree 6

$$Y = 39.077 + 0.0307X - 3E\text{-}05X^2 + 1E\text{-}08X^3 - 2E\text{-}12X^4 + 2E\text{-}16X^5 - 5E\text{-}21X^6$$
with $R^2 = 0.22$

*Hyperbolic:*
$$Y = 101.31 X^{-0.1102}$$
with $R^2 = 0.04$

*Exponential:*
$$Y = 44.846 \, e^{-2E\text{-}05 \, X}$$
with $R^2 = 0.04$

If one compares the coefficient of determination results between annual Dst sum and annual Dst mean the differences are practically non-existing.

We use the coefficient of determination as an indicator for the goodness of fit of the curve to the data. Clearly the largest coefficient of determination $R^2 = 0.35$ was obtained using SSN as the independent variable and a polynomial of degree 6 as formula. From the more direct indices the largest coefficient of determination was obtained using the annual Dst mean with $R^2 = 0.23$ and again a polynomial of degree 6. Very close lies the coefficient of determination using the Annual no of days with Dst $\leq -40$, $R^2 = 0.22$ It is perhaps surprising to see better fitting when we use the SSN which is not a direct index. The explanation for this is mathematical convenience: This is a physics vs maths game, physics prefers Dst as being a more direct index for our purpose yet maths (statistics) prefer SSN which has a smoother graph as a function of time (we discuss this issue in greater detail in the last section "Conclusions").

## *II. Local Indices*

The long term solar activity indices used in the previous sub-section were defined using planetary averages data like the Ap and Dst indices data. In this sub-section we use some local geomagnetic data from the Geomagnetic Observatory of Penteli (GOP for short, its I.M.O. code is PEG) which is located about 30 km North East of Athens

http://www.magstation.gr/geomagnetism/wiki/Main

GOP belongs to IGME, the (Greek) Institute of Geology and Mineral Exploration

http://www.igme.gr/portal/page?_pageid=33,56803&_dad=portal&_schema=PORTAL

This is the only geomagnetic station existing in Greece. It became a member of IMO (Intermagnet group/network) on the 8[th] of Nov 2010

http://www.magstation.gr/geomagnetism/wiki/News

For the period we are interested in, namely 1989-2010, GOP could only provide the 3h K-index data. Using the data provided by GOP, as a long term (direct, local) solar activity index we defined the annual mean of the daily sum of the 8 3h K-indices measured by GOP:

For instance Penteli gave us the 3 h K-index for all days from 1989-2010. As an example, for the 1[st] Jan 2010 we were given the following data

Table 4: Definition of the annual mean of the daily sum of the 3h K-indices (Penteli)

| day | K-index | | | | | | | | |
|---|---|---|---|---|---|---|---|---|---|
| | 00.00-03.00 | 03.00-06.00 | 06.00-09.00 | 09.00-12.00 | 12.00-15.00 | 15.00-18.00 | 18.00-21.00 | 21.00-24.00 | **Daily K-index SUM** |
| 1[st] Jan 2010 | 0 | 0 | 0 | 0 | 0 | 1 | 0 | 2 | **3** |
| 2[nd] Jan 2010 | 0 | 0 | 1 | 1 | 1 | 0 | 0 | 0 | **3** |
| Etc… | | | | | | | | | |

Thus we computed the annual mean of the daily K-index sum and this is what we used as a long term (local) solar activity index.

Table 5 below gives the annual mean of the daily K-index sums from GOP:

Table 5: Annual Average of the daily K-index sum (GOP)

| No | Year | Annual average of the daily K-index sum (GOP) |
|---|---|---|
| 1 | 1989 | 19.58 |
| 2 | 1990 | 18.27 |
| 3 | 1991 | 20.14 |
| 4 | 1992 | 18.91 |
| 5 | 1993 | 17.33 |
| 6 | 1994 | 17.68 |
| 7 | 1995 | 14.75 |
| 8 | 1996 | 14.63 |
| 9 | 1997 | 13.11 |
| 10 | 1998 | 15.39 |
| 11 | 1999 | 16.94 |
| 12 | 2000 | 18.59 |
| 13 | 2001 | 17.32 |
| 14 | 2002 | 17.47 |
| 15 | 2003 | 21.98 |
| 16 | 2004 | 16.88 |
| 17 | 2005 | 17.07 |
| 18 | 2006 | 14.24 |
| 19 | 2007 | 13.26 |
| 20 | 2008 | 13.12 |
| 21 | 2009 | 10.64 |
| 22 | 2010 | 12.52 |

In the figure 4 below we give the graphs of SSN and (scaled) Annual mean of daily K-index sum from Penteli vs Time. One can see that it follows the general pattern of SSN, periodicity, two picks and 2 valleys

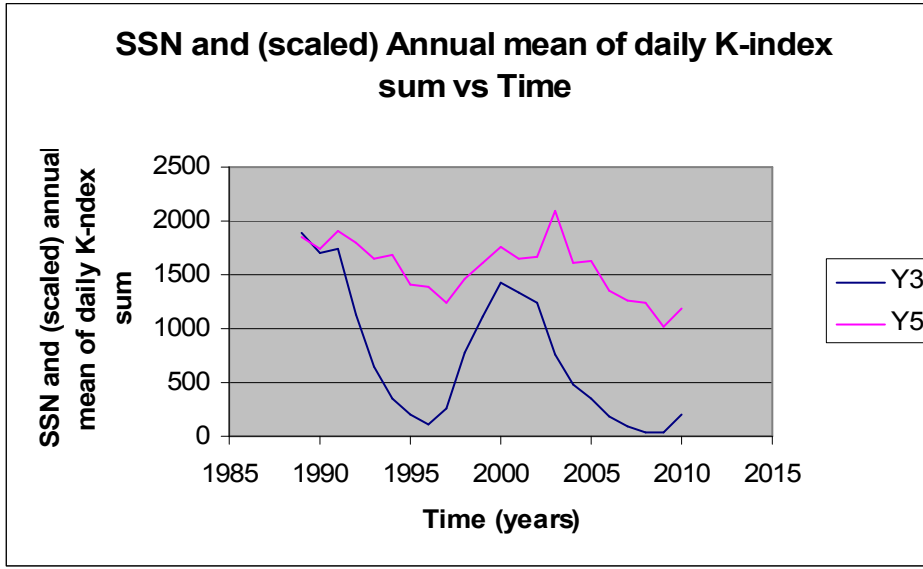

Figure 4: SSN (blue) and scaled (multiplied by a factor of 95) Annual mean of the daily K-index sum from Penteli (purple) vs Time graph.

Next we run the same regression model as we did in the previous subsection (both linear and non-linear) and we write down our results:

*Linear:*
$$Y = -0.5098\ X + 65.602$$
with $R^2 = 0.13$

*Logarithmic:*
$$Y = -23.511\ \ln X + 106.26$$
with $R^2 = 0.13$

*Polynomial:* Among polynomials in X with degrees from 2 up to 6, the largest coefficient of determination was given using a polynomial of degree 6

$$Y = 89181 - 34006X + 5330.3X^2 - 439.69X^3 + 20.146X^4 - 0.4865X^5 + 0.0048X^6$$

with $R^2 = 0.23$

*Hyperbolic:*
$$Y = 208.03\ X^{-0.5995}$$
with $R^2 = 0.13$

*Exponential:*
$$Y = 73.082\ e^{-0.0379\ X}$$
with $R^2 = 0.13$

It is clear that the best fitting curve is the polynomial of degree 6 with a coefficient of determination equal to $R^2 = 0.23$. In the following graph we give the scatter plot along with the polynomial curve. We summarize our regression results using Penteli data below:

Table 6: Summary of Regression results (local geomagnetic data from Penteli)

| Index | Function | Equation | $R^2$ |
|---|---|---|---|
| Annual average of the daily K-index sum (Penteli) | linear | $Y = -0.5098 X + 65.602$ | 0.13 |
| | ln | $Y = -23.511 \ln X + 106.26$ | 0.13 |
| | poly | $Y = 89181 - 34006X + 5330.3X^2 - 439.69X^3 + 20.146X^4 - 0.4865X^5 + 0.0048X^6$ | 0.23 |
| | hyp | $Y = 208.03 X^{-0.5995}$ | 0.13 |
| | exp | $Y = 73.082 e^{-0.0379 X}$ | 0.13 |
| | | | |

Now in order to be as mathematically rigorous as possible, for this "best fitting curve" (polynomial of degree 6), we have to check some assumptions regarding non-linear regression and the ordinary least squares method used to estimate the relevant parameters. The basic assumptions we have to check is homoscedasticity and normality of the residuals (we assume weak exogeneity and there is no question about multicollinearity since essentially we use only one independent variable). If these tests fail that would mean that the ordinary least squares method used to estimate the parameters of the order 6 polynomial curve should be replaced by a more elaborate method (like weighted least squares or perhaps generalised least squares or iteratively weighted least squares).

The residuals are given in table 7 below:

Table 7: Residuals of polynomial of order 6 curve (Annual average of daily K-index sum vs TFN)

| No | Residuals |
|---|---|
| 1 | -2.8576787 |
| 2 | -12.2688403 |
| 3 | 2.1125095 |
| 4 | 7.8219997 |
| 5 | 23.4029636 |
| 6 | 5.4443677 |
| 7 | 5.5248021 |
| 8 | -13.9868961 |
| 9 | 2.5054937 |
| 10 | -5.6723138 |
| 11 | 0.7507142 |
| 12 | -3.7880477 |
| 13 | -8.6053077 |
| 14 | -13.5245735 |
| 15 | -0.1795239 |
| 16 | -14.4118189 |
| 17 | 20.0451768 |
| 18 | 9.0543094 |
| 19 | 9.0229855 |
| 20 | -7.5998873 |
| 21 | 0.2454871 |
| 22 | -3.0359214 |

In order to check normality we perform the Shapiro Wilk test (Shapiro & Wilk 1965) for the residuals. Essentially we test the null hypothesis that the residuals are normally distributed against the alternative that they are not. At the a = 0.05 confidence level, in our case the W statistic equals 0.93 and since the threshold equals 0.91, we cannot reject the null hypothesis, hence the null hypothesis is accepted and thus the residuals come from a normal population.

In order to check homoscedasticity we perform the Bartlett test of homogeneity of variances (see Snedecor & Cochran 1989) for the residuals. We divide the residuals in 2 groups, the first group consists of the first 11 and the second group consists of the remaining 11 residuals (22 in total). We compute the variances of the 2 samples, the first one equals 103.4, the second one equals 108.3. The Bartlett test is essentially the test of the null hypothesis that the variances of the two samples are equal against the alternative that they are not. The relevant test statistic, Bartlett's K square follows the chi square distribution. In our case we have 2 samples hence 1 degree of freedom. At the a = 0.05 confidence level the critical value is 0.946. The Bartlett's K square we compute is 0,00861. Since this is not larger than the critical value we cannot reject the null hypothesis that the variances are equal, thus we accept the null hypothesis that the variances are equal and hence the residuals (errors) are homoscedastic.

# 5. Short term (immediate) effects

For an assessment of the short term effects of solar activity onto large transformers we follow what we described in section 2 above.
A possible simple "cumulative short term effect index" would be the average (over the 22 years) of the fraction

(failures days with Ap≥100) / (No of days with Ap≥100)

And similarly for the Dst index.

Alternatively we could take the sum of failure days with Ap≥100 from 1989-2010 and divide it with the total sum of days with Ap≥100 again from 1989-2010 (and similarly for the Dst index). This will give for Ap the fraction

$$11/43 \approx 0.26$$
(normalized fraction $6/43 \approx 0.14$)

and similarly for the Dst one computes

$$13/82 \approx 0.16$$

In order to be able to interpret the above numbers *as the (empirical) probability (or experimental probability or relative frequency) to have a failure during a stormy day* the quotients have to be normalized. A simple way to do that would be to consider only one failure for each stormy day (there may be more than one) and to consider strictly failures occurring during stormy days alone (and not failures during "around" stormy days). This shows that Ap is perhaps a "more indicative" index for immediate-short term effects in Greece.
Similar simple indices can be calculated and they can be interpreted (after proper normalization) as the experimental (a posteriori) probability to have a failure during a stormy day plus or minus 3 days (and similarly for the 7 days case).

Next we present briefly some indicative cases along with some comments. We focus on cases with Ap ≥100 or Dst ≤-100 or both:

*1989*
On 21.10.1989 the planetary Ap mean was 146 and the Dst planetary mean was -191. In an 150kV transformer in Central Greece, the Buchholz relay of the tap changer was triggered. The cause was recorded as "unknown". This incident is a very possible immediate effect of solar activity (in particular GIC's) on a transformer.

In 1989 there was the Quebec black out in Canada due to the solar super storm on 13-14 March 1989. There were no (immediate) recorded failures in the Greek power grid caused by this solar storm.

*1990*
On 30.03.1990 where the planetary Ap mean was 69 and the Dst planetary mean was -106. In an 150 kV transformer in NE Greece there was a tap changer malfunction; the tap changer was replaced.

*1991*
3 Case studies for immediate effects:

On 24.03.1991 where the planetary Ap mean was 161 and the Dst planetary mean was -75. In an 150 kV transformer in N of Athens there was a tap changer malfunction which was repaired.

On 05.06.1991 where the planetary Ap mean was 196 and the Dst planetary mean was -147. In an 150 kV transformer in N of Athens there was a tap changer malfunction which was repaired.

On 29.10.1991 where the planetary Ap mean was 128 and the Dst planetary mean was -173. In an 150 kV transformer in NE Greece the differential protection relay was triggered; also some oil leakage was observed. The tap changer was replaced, insulator replaced.

1991 was a very active year, there were 11 days in total with Ap≥100: 24, 25, 26 March and there was a reported failure on the $24^{th}$ and one on the $20^{th}$. In June on $5^{th}$, $10^{th}$ and $13^{th}$ Ap was larger or equal than 100 and we had 1 reported failure on the $5^{th}$ and 4 more in between a week of the stormy days. In July the Ap was larger or equal than 100 on the $9^{th}$ and $13^{th}$ and we had 2 failures on the $7^{th}$ and on the $6^{th}$ (3 days before the $9^{th}$). Finally Ap was larger or equal than 100 on 29.10, 01.11 and 09.11. We had a reported failure on 29.10, one reported failure on the 25.10, one on 8.11 and one on 14.11.

*1992*
On 27.02.1992 where the planetary Ap mean was 60 and the Dst planetary mean was -101. In an 150 kV transformer in N Greece the differential protection and Buchholz relays were triggered and the failure was repaired.

*1993*
On 13.09.1993 where the planetary Ap mean was 91 and the Dst planetary mean was -109. In an 150 kV transformer in Central Greece there was a tap changer malfunction which was repaired.

*1995*
On 07.04.1995 where the planetary Ap mean was 100 and the Dst planetary mean was -82. In two 150 kV transformers, one in the island of Corfu and another in N Greece there were reported failures: In the first case there was inability to switch from 15 kV to 21 kV and in the second the Buchholz relay was triggered and oil leakage was observed. In both cases the failures were repaired.

*2000*

On 18.09.2000 where the planetary Ap mean was 70 and the Dst planetary mean was -104. In one 150 kV transformer in N Greece there was oil leakage due to an unknown reason. The failure was repaired.

*2001*

On 01.04.2001 where the planetary Ap mean was 38 and the Dst planetary mean was -137. In one 150 kV transformer in N Greece there was a failure characterized as "mechanical", switch transmission axis. The recorded cause for the failure was "aging". The failure was repaired.

*2005*

1 possible Case study for an immediate effect: On 11.09.2005 where the planetary Ap mean was 101 and the Dst planetary mean was -84. In one 150 kV transformer close to Athens there was a failure in a casing gasket which was replaced. Also oil leakage was observed. The failure was due to "manufacture defect" according to the failure report.

**6. Conclusions**
In this work we tried to provide evidence that solar activity affects large transformers (400 kV and 150 kV) in Greece. Effects are divided into short term (immediate) and long term effects. The common belief in PPC for many years (in fact since its founding in 1950's) was that there are no such effects because Greece is a mid-latitude country (latitude approximately $35^o – 41^o$ North). The second goal was to provide a method to assess solar activity long term effects on large transformers by using various analytic and statistical methods.

Let us start with the short term (immediate) effects: We used transformer failures data for the period 1989-2010 provided by IPTO (ADMIE), the Greek independent power transmission operator, a fully subsidiary company of PPC (the Public Power Corporation of Greece). After some data clearance (in order to get rid of obviously irrelevant to GICs failures), we ended up with a list of transformer failures which could be related to GICs. This task had some difficulties since PPC personnel are not trained on GIC's (like say their Scandinavian co-workers where for example technicians have to fill in specific repairing forms in the shape of a questionnaire, thus helping to identify GIC related failures easier). Nonetheless, having our set of transformer failure data at hand we juxtaposed it with solar activity data and we saw classic cases of GIC related failures like triggering of protecting relays, oil leakage due to insulation destruction etc, appearing on days of elevated solar activity (say Ap index larger than 100). We obtained clearance to publicise some of these cases and briefly analysed them in section 5.
However there are also rapid variations of the terrestrial magnetic field a few days prior to a solar storm and for this reason we extended our research not only on stormy days but also on short periods around a stormy day: We chose two additional time scales, ±3 days and ±7 days around a stormy day. The plus or minus 3 days time scale was chosen in order to include effects due to sudden commencement of a storm -- primarily for the "minus" part. We also observed that there is an important increase in failures during days +4, +5, +6 and +7 from a stormy day and thus we decided to include the ±7 days time scale as well. Taking a closer look of the data we saw evidence that this second scale captures failures triggered shortly after a storm and reveals "weaknesses" of the transformers like bad maintenance, etc. Hence during the period of interest (1989-2010), there were 43 stormy days (namely days with Ap≥100) and we observed 19 failures occurring during a stormy day plus or minus 3 days and 51 failures occurring during a stormy day plus or minus 7 days.

As an attempt to assess in a cumulative manner the probability to have a transformer failure somewhere in Greece during a day of high solar activity we simply counted the number of stormy days (namely days with Ap ≥ 100 or Dst ≤ -100) and then we counted in how many of them we observed a transformer failure and by division we got an idea of a probability (or relative frequency) to have a transformer failure during a stormy day (or around a stormy day).

For the long term effects: We had to define some new long term solar activity indices and then we tried to quantify the long term effects using statistical and analytic techniques. Long term solar activity indices have been defined using both global (namely planetary) short term indices data (like Ap and Dst data) and local (from the Penteli Magnetic Observatory near Athens) short term indices data (like the 3h K-index data). In particular we defined and used 6 long term indices, 5 using global data and 1 using local data: The former include the annual number of sunspots (SSN), the annual Ap mean, the annual Dst mean, the annual number of disturbed days, namely days with Ap ≥ 40 and the annual number of days with Dst ≤ - 40. The only local geomagnetic data available for the period of interest (1989-2010) was the 3 hour K-index provided by the Geomagnetic Observatory of Penteli near Athens. Using these K-index data we defined the annual mean of the daily K-index sum which we used as a long term local solar activity index.

The first important fact revealed from our analysis, concerning the long term effects of solar activity onto large transformers is shown in Fig 3 where we plotted both SSN (annual sunspot number) and TFN (annual transformer failure number) as functions of time in the same graph: In Fig 3 we observe two striking (at least for PPC technicians) facts:

1.Apart from noise, failures do show some periodicity similar to the periodicity of solar activity (the solar cycles).
2. The 2 graphs seem to have a "phase delay" roughly half a solar cycle (more precisely 4-5 years).

Let us make some comments on these facts: This rather revealing graph indicates that the annual number of failures generally follows the solar activity pattern, it has a periodicity similar to the 11 year solar cycle but it has a "phase delay": During solar cycle 22 (1986-1996) the max number of sunspots was observed in 1989 and the max number of failures was observed in 1993 (4 years later). During solar cycle 23 (1996-2008) the max number of sunspots was observed in 2000 and the max number of failures was observed in 2006 (6 years later). So by average the phase delay is about 5 years later, roughly half a solar cycle. This can be explained by the fact that GIC's in Greece are existing but smaller compared to say Scandinavian countries and cumulative effects dominate; there is a building up phenomenon related say to erosion of transformer components, say insulators: Failures occur after some threshold is crossed and since GIC's in Greece (latitude approximately $35^o$ - $41^o$ N) are generally smaller, this threshold is crossed later (the erosion of components is not so fast). So the picture is that as latitude increases, this phase delay should decrease, namely in say Scandinavian countries one would expect very little or no phase delay at all (well, one has to be more careful here since failures depend on ground conductivity, grid topology etc, the above conjecture assumes that all other "variables" are the same, say if one could move Greece and its grid in the position of Finland). Note however that if one uses another more direct long term earth affecting solar activity index, this time difference gets shorter: for example the max Ap mean, the max Dst mean, the max of annual number of days with Ap ≥ 40 and the max of annual number of days with Dst ≤ - 40 for solar cycle 22 they all occurred in 1991 --and not in 1989 as sunspots did-- and in this case the phase delay was 2 years and not 4. For solar cycle 23, the max Ap mean, the max Dst mean, the max of annual number of days with Ap ≥ 40 all occurred in 2003 (yet the max of annual number of days with Dst ≤ - 40 occurred in 2002) and not in 2000 as sunspots did and in this case the phase delay was 3 years and not 6. Fig 3 along with the above discussion also explains the negative sign of the correlation coefficient between TFN and SSN (linear regression approximation).

Then we applied various statistical techniques, regression and correlation, using a number of function models: linear, polynomial (of various degrees), logarithmic, exponential and hyperbolic and we computed the corresponding equations (coefficients) and coefficient of determination. We gathered all these regression results in Table 3 (for the global long tern solar activity indices) and in Table 6 (section 4.II) we presented the regression results for the local long term solar activity index (data from Penteli).

Now let us comment on the regression results: The best results (larger coefficient of determination) in all cases were given by polynomials and we provide a theoretic explanation for this in an Appendix, based on the Stone-Weierstrass Theorem from functional analysis.
Another interesting point here is that the highest coefficient of determination (0.35) was obtained using SSN (along with a polynomial of $6^{th}$ degree). This is perhaps a little surprising since SSN is a global and indirect (for our purposes) index. One might expect that the best results would have been obtained with the local K-index data (direct and local index). The explanation for this fact is, we believe, that the SSN curve is smoother and it complies better with the mathematical requirements of regression (see also below). However, in the Appendix where we push polynomial regression to the limit and we use more sophisticated mathematical tools, the best result ($R^2 = 0.88$) is obtained by the "direct" index annual no of days with Dst ≤ -40.

We would also like to make a comment on the use of the specific data to define our long term indices: The reason for choosing to rely primarily (but not exclusively) on Ap and Dst data throughout this work (which are planetary averages) is this: There is only one geomagnetic station in Greece, the Penteli Magnetic Observatory. It is understandable that the "slope" $\Delta B/\Delta t$ with $\Delta t \sim 1min$ (or the geoelectric field) would be more physically relevant short term geomagnetic local indices.
These local data ($\Delta B/\Delta t$ with $\Delta t \sim 1min$) are not publicly available from the magnetic observatory of Penteli; in fact we are not even sure if they exist for the period of interest (1989-2010) since Penteli became a member of the World Data Centre for Geomagnetism from late 2010 onwards.

As far as the geoelectric field is concerned, we are not aware of any institution in Greece making such measurements.

We would like to mention however that even if these local data were available, one has to bear in mind that the final step would be to use them to define a long term index (on an annual basis). The most obvious choice would be to compute some annual average. It is probable that the "sensitivity" of these short term local indices might be lost when we pass on from short term to the long term. For example if one computed the annual average of the 1 min slope $\Delta B/\Delta t$, there are half a million minutes in a year and thus this long term index (annual one minute variation of the magnetic field) might be almost constant throughout the 11 year solar cycle. The next step would be to apply statistical regression techniques. These mathematical methods rely on certain mathematical theorems which in turn are based on certain assumptions (the most important ones are that the variables are non-stochastic, weak exogeneity holds, homoscedasticity holds, normality of residuals holds, there is no self-correlation of the residuals, the covariance between the variable values and the error term tends to zero etc—we checked some of them for one specific case, the annual average of the daily K-index sum from Penteli); it is not guaranteed that the most physically relevant indices are those which satisfy the mathematical assumptions the most and hence give the largest coefficient of determination (this is the meaning of the phrase in the text that "this is a physics vs maths game" and that "the SSN time curve has a smoother graph"). As we can see from our results, for example SSN is an indirect earth affecting solar activity index yet nonetheless gives a satisfactory coefficient of determination—perhaps a rather surprisingly satisfactory coefficient of determination. Hence our approach to use statistics and the coefficient of determination as a criterion (which, let us recall does not necessarily mean causation but it does mean predictability according to one of the golden rules in statistics) gives in some sense the optimal result between basically three pairs of opposite parameters: short term vs long term, physical relevance vs mathematical convenience, global vs local geomagnetic data. Needless to say if we had these data we would have run our models for them too (which means that other countries for which these local data are available should try them).

It is straight forward that the physical quantity which is the most physically relevant for our study are the GIC's. The ideal would be to have local measurements for GIC's, define some long term index from them and then apply our statistical models between GIC's and TFN. Yet these currents have never been measured directly anywhere in Greece during the past and to the best of our knowledge there is no program to do so in the future. It is well known however that there is a good linear approximation between GIC's and the geoelectric field (see for example Pulkinen 2003, Lindahl 2003, Viljanen et al. 2004, Pulkinen et al. 2008, Pulkinen et al., 2010, Pulkinen et al. 2012), given by the following equation

$$GIC = a\, E_x + b\, E_y,$$

where (a, b) are system specific parameters typically in the range of 0-200 (in A.km/V) and $E_x$ and $E_y$ are the horizontal components of the geoelectric field (y-axis in the east-west direction with the positive y-semi-axis pointing to the east and the x-axis in the south-north direction with the positive x-semi-axis pointing to the north). If no transformer data is known then can give a good approximation of the sum of GIC flowing over all phases of the transformer. Unfortunately, to the best of our knowledge, there are no measurements of the geoelectric field anywhere in Greece either.

One can compute the geoelectric field from the geomagnetic field (horizontal components $B_x$, $B_y$) using the relations suggested in the literature given above (Pulkinen 2003, Lindahl 2003, Viljanen et al. 2004, Pulkinen et al. 2008, Pulkinen et al., 2010, Pulkinen et al. 2012):

$$E_x(\omega) = [Z(\omega)\, B_y(\omega)]/\mu_0$$

and

$$E_y(\omega) = [Z(\omega)\, B_x(\omega)]/\mu_0$$

where $\mu_0$ is the permeability of the vacuum, $\omega$ is the angular frequency of the fluctuating field (about 0.01-0.1 Hz) and $Z(\omega)$ denotes the surface impedance which depends on the ground conductivity. These equations are derived from Faraday's Law making various simplifications. Again, to the best of our knowledge there are no measurements of the surface impedance in Greece.

So the key thing is that we believe we did our best with the data available.

**Acknowledgements:** The author wishes to thank the following people for providing data and for useful discussions: Athanasios Tzortzis (TRSC, PPC Greece), Constantine Mavromatos (IPTO, ADMIE Greece), P. Kapiris, P. Tsailas and G. Filippopoulos (GOP and IGME Greece), Mike Hapgood (RAL, Oxford, UK), Sarah James (UKSSDC, RAL Space, Oxford UK), Hans-Joachim Linthe (Helmholtz Centre Potsdam Germany), Brent Gordon, Robert Rutledge and William Murtagh (NOAA, Space Weather Prediction Centre, Boulder Colorado USA) and Masahito Nose (World Data Centre for Geomagnetism, Kyoto Japan).

# Mathematical Appendix: Polynomial regression taken to the limit

We think that it might be helpful for the reader to clarify some mathematical aspects of our work in the current appendix.

We start by stating our problem in mathematical terms: let X denote the independent variable which is some long term annual (or monthly) solar activity index and let Y denote the dependent variable which is the annual (or monthly) transformer failure number. The mathematical problem we faced was this: Approximate the true but unknown function Y = F(X) relating the 2 variables X and Y above using another "calculable" function Y = f(X) (all functions are assumed real with real variables).

We make the perfectly reasonable (by all aspects) ansatz that F is continuous. Then we can use the famous Stone-Weierstrass theorem (see Rudin 1976, Glicksberg 1962 and Jeffreys 1988) from functional analysis which states that:

*"Let K be a compact metric space and let us denote by C(K;R) the Banach algebra of continuous real valued functions defined on K, equipped with the sup-norm. Let A be a unital sub-algebra of C(K;R) which separates points of K. Then A is dense in C(K;R)".*

(Further generalisations exist for complex algebras, for locally compact spaces, the Bishop-Silov theorem etc).

We can apply the aforementioned theorem in our case, making the following choices: As K we pick any closed interval [a,b] in the real numbers and as A we chose the simplest available unital sub-algebra of C([a,b],R) which separates points in [a,b]: This is the algebra of ordinary polynomials. Then the theorem states that polynomials are (uniformly) dense in C([a,b],R) with respect to the sup-norm. In plain English: Any continuous real function defined on a closed and bounded interval can be uniformly approximated (up to any degree of accuracy) using polynomials (of sufficiently large degree). So our problem has been solved in principle (all spaces are in our case compact and hence uniform continuity implies continuity). This fact explains in principle why polynomial regression gave the best results in all cases (that is for all long term indices defined).

Now the additional difficulty in our case at hand is that we do not know F explicitly. Thus in order to compute specific examples and find a measure of how good our approximation is, we can use the data points available directly (instead of an explicit expression for F and the sup-norm) and then rely on statistical methods --like the least squares method (and its ramifications). For the calculations (which are rather tedious) we relied primarily on R (special open source computer language for statistical computations—this is what most statistics researchers in academia use-- and MS Excel to a lesser extend).

As calculable examples then, for each one of the five long term (earth affecting) solar activity indices, namely annual Ap mean, Annual Dst mean, annual number of days with Ap ≥ 40, annual number of days with Dst ≤ -40 and annual mean of the daily K-index sum from the Penteli observatory—local data (we chose the annual basis for computations of examples), we have run the following linear and non-linear regression models Y = f(X) with f being one of the following type of functions:
- linear,
- hyperbolic,
- logarithmic,

- exponential,
- polynomial of degree 1 up to 6.

We used the coefficient of determination $R^2$ as a measure of the "goodness of fit" of our data in the scatter plot with the predicted equation. The best result (largest $R^2$) for all long term indices was obtained using polynomial regression. Clearly the deep reason is the Stone-Weierstrass theorem. However we thought that one could try two further things to improve the fit:

(I.) Try to push polynomial regression to the limit, namely increase the polynomial degree up to the maximum value for which the least squares method can be successfully applied (i.e. the coefficients can be determined—namely one does not have an over-constrained or over-determined system, that the coefficients are unique and without singularities, Runge's phenomenon appearing etc).

(II.) Try to use another, more elaborate unital subalgebra of $C([a,b],R)$ which separates points in $[a,b]$, like the algebra of orthogonal polynomials instead of the algebra of ordinary (raw) polynomials (see for example Zois 2009 and Zois 2010). We thus have the following results (we briefly include some indicative R language commands and the corresponding summary results—after the summary command, the second column gives the coefficients of the corresponding power written in italics, as references for the orthogonal polynomials used in R see Chambers & Hastie 1992 and Kennedy & Gentle 1980):

**1. Polynomial regression between failures (TFN, Y) and sunspots (SSN, X)**

```
> xssn <- c(1893, 1707, 1749, 1134, 657, 358, 210, 103, 258, 770, 1118, 1434, 1331, 1249, 763, 485, 357, 182, 90, 34, 37, 198)
> ytfn <- c(35, 25, 38, 46, 60, 42, 55, 36, 48, 40, 38, 34, 28, 23, 34, 23, 57, 60, 56, 38, 49,35)
> sample1 <- data.frame(xssn, ytfn)
> sample1
   xssn ytfn
1  1893  35
2  1707  25
3  1749  38
4  1134  46
5   657  60
6   358  42
7   210  55
8   103  36
9   258  48
10  770  40
11 1118  38
12 1434  34
13 1331  28
14 1249  23
15  763  34
16  485  23
17  357  57
18  182  60
```

19  90  56
20  34  38
21  37  49
22 198  35

*A. Ordinary (Raw) polynomials*

> fit1 <- lm(sample1$ytfn ~ poly(sample1$xssn, 12, raw=TRUE))
> summary(fit1)

Call:
lm(formula = sample1$ytfn ~ poly(sample1$xssn, 12, raw = TRUE))

Residuals:
    Min      1Q  Median      3Q     Max
-12.6938 -5.3335 -0.1291  5.1719 15.8137

Coefficients (second column in italics, coefficient of the corresponding monomial power):
                                    Estimate Std. Error t value Pr(>|t|)
(Intercept)                        -7.473e+01  7.311e+01  -1.022   0.3334
poly(sample1$xssn, 12, raw = TRUE)1   *6.231e+00*  3.607e+00   1.728   0.1181
poly(sample1$xssn, 12, raw = TRUE)2  *-1.098e-01*  6.048e-02  -1.816   0.1028
poly(sample1$xssn, 12, raw = TRUE)3   *9.341e-04*  4.926e-04   1.896   0.0904 .
poly(sample1$xssn, 12, raw = TRUE)4  *-4.467e-06*  2.278e-06  -1.961   0.0815 .
poly(sample1$xssn, 12, raw = TRUE)5   *1.311e-08*  6.521e-09   2.011   0.0753 .
poly(sample1$xssn, 12, raw = TRUE)6  *-2.490e-11*  1.215e-11  -2.049   0.0707 .
poly(sample1$xssn, 12, raw = TRUE)7   *3.150e-14*  1.515e-14   2.080   0.0673 .
poly(sample1$xssn, 12, raw = TRUE)8  *-2.675e-17*  1.271e-17  -2.105   0.0646 .
poly(sample1$xssn, 12, raw = TRUE)9   *1.507e-20*  7.085e-21   2.126   0.0624 .
poly(sample1$xssn, 12, raw = TRUE)10 *-5.394e-24*  2.515e-24  -2.145   0.0605 .
poly(sample1$xssn, 12, raw = TRUE)11  *1.111e-27*  5.140e-28   2.162   0.0589 .
poly(sample1$xssn, 12, raw = TRUE)12 *-1.002e-31*  4.604e-32  -2.177   0.0575 .
---
Signif. codes:  0 '***' 0.001 '**' 0.01 '*' 0.05 '.' 0.1 ' ' 1

Residual standard error: 11.25 on 9 degrees of freedom
Multiple R-squared: 0.5982,    Adjusted R-squared: 0.06256
F-statistic: 1.117 on 12 and 9 DF,  p-value: 0.4434

Raw polynomials with degrees higher than 12 give singularities (coefficients cannot be computed).

*B. Orthogonal polynomials*

With orthogonal polynomials one can go up to degree 17 (for higher degrees the least squares method does not give a solution, one has an over-constrained system, the degree must be less than the number of unique points):

> fit2 <- lm(sample1$ytfn ~ poly(sample1$xssn, 17))

> summary(fit2)

Call:
lm(formula = sample1$ytfn ~ poly(sample1$xssn, 17))

Residuals:
```
         1          2          3          4          5          6          7
 8.983e-06  7.529e-04 -4.347e-04  1.651e+00  8.671e-01 -5.413e+00  8.930e+00
         8          9         10         11         12         13         14
-4.292e+00 -1.975e+00  8.197e+00 -1.525e+00 -1.542e-02  1.085e-01 -2.765e-01
        15         16         17         18         19         20         21
-8.782e+00 -5.001e-01  6.238e+00  6.014e+00  3.954e+00  2.097e+00 -2.529e+00
        22
-1.275e+01
```

Coefficients:
```
                         Estimate Std. Error t value Pr(>|t|)
(Intercept)               40.9091     2.4905  16.426 8.04e-05 ***
poly(sample1$xssn, 17)1  -27.6923    11.6815  -2.371   0.0768 .
poly(sample1$xssn, 17)2    2.6679    11.6815   0.228   0.8305
poly(sample1$xssn, 17)3    9.6013    11.6815   0.822   0.4573
poly(sample1$xssn, 17)4    1.2141    11.6815   0.104   0.9222
poly(sample1$xssn, 17)5    4.4720    11.6815   0.383   0.7213
poly(sample1$xssn, 17)6  -10.1385    11.6815  -0.868   0.4344
poly(sample1$xssn, 17)7    0.5499    11.6815   0.047   0.9647
poly(sample1$xssn, 17)8    4.2829    11.6815   0.367   0.7325
poly(sample1$xssn, 17)9   -4.6379    11.6815  -0.397   0.7116
poly(sample1$xssn, 17)10  -4.0774    11.6815  -0.349   0.7447
poly(sample1$xssn, 17)11  -6.9838    11.6815  -0.598   0.5821
poly(sample1$xssn, 17)12 -24.4810    11.6815  -2.096   0.1042
poly(sample1$xssn, 17)13   4.8624    11.6815   0.416   0.6986
poly(sample1$xssn, 17)14  13.5348    11.6815   1.159   0.3111
poly(sample1$xssn, 17)15  16.8430    11.6815   1.442   0.2228
poly(sample1$xssn, 17)16  -3.9135    11.6815  -0.335   0.7544
poly(sample1$xssn, 17)17  -9.3193    11.6815  -0.798   0.4697
```
---
Signif. codes:  0 '***' 0.001 '**' 0.01 '*' 0.05 '.' 0.1 ' ' 1

Residual standard error: 11.68 on 4 degrees of freedom
Multiple R-squared: 0.8074,   Adjusted R-squared: -0.01122
F-statistic: 0.9863 on 17 and 4 DF,  p-value: 0.5721

2. **Polynomial regression between failures (TFN, Y) and annual Ap mean (AAPM, X)**

*A. Raw Polynomials*

Coefficients:
```
                 Estimate Std. Error t value Pr(>|t|)
(Intercept)     -1.424e+04  6.694e+04  -0.213    0.835
```

```
poly(sample2$xaapm, 10, raw = TRUE)1    1.595e+04  6.706e+04   0.238    0.816
poly(sample2$xaapm, 10, raw = TRUE)2   -7.558e+03  2.893e+04  -0.261    0.799
poly(sample2$xaapm, 10, raw = TRUE)3    2.006e+03  7.105e+03   0.282    0.783
poly(sample2$xaapm, 10, raw = TRUE)4   -3.313e+02  1.103e+03  -0.300    0.769
poly(sample2$xaapm, 10, raw = TRUE)5    3.574e+01  1.133e+02   0.315    0.758
poly(sample2$xaapm, 10, raw = TRUE)6   -2.562e+00  7.821e+00  -0.328    0.749
poly(sample2$xaapm, 10, raw = TRUE)7    1.210e-01  3.591e-01   0.337    0.742
poly(sample2$xaapm, 10, raw = TRUE)8   -3.619e-03  1.052e-02  -0.344    0.737
poly(sample2$xaapm, 10, raw = TRUE)9    6.210e-05  1.777e-04   0.349    0.733
poly(sample2$xaapm, 10, raw = TRUE)10  -4.657e-07  1.319e-06  -0.353    0.731
```

Multiple R-squared: 0.2187,    Adjusted R-squared: -0.4917

Raw polynomials of degree higher than 10 give singularities and the coefficients cannot be computed.

*B. Orthogonal polynomials*

```
Coefficients:
                              Estimate Std. Error t value Pr(>|t|)
(Intercept)                    40.9091     3.1701  12.905 4.14e-07 ***
poly(sample2$xaapm, 12)1      -14.8621    14.8692  -1.000    0.344
poly(sample2$xaapm, 12)2        4.0447    14.8692   0.272    0.792
poly(sample2$xaapm, 12)3        0.6927    14.8692   0.047    0.964
poly(sample2$xaapm, 12)4       -2.9508    14.8692  -0.198    0.847
poly(sample2$xaapm, 12)5        3.0785    14.8692   0.207    0.841
poly(sample2$xaapm, 12)6        9.5484    14.8692   0.642    0.537
poly(sample2$xaapm, 12)7      -10.6245    14.8692  -0.715    0.493
poly(sample2$xaapm, 12)8       11.3732    14.8692   0.765    0.464
poly(sample2$xaapm, 12)9       -2.2829    14.8692  -0.154    0.881
poly(sample2$xaapm, 12)10      -5.0101    14.8692  -0.337    0.744
poly(sample2$xaapm, 12)11      -0.5810    14.8692  -0.039    0.970
poly(sample2$xaapm, 12)12     -14.9675    14.8692  -1.007    0.340
---
```

Multiple R-squared: 0.2978,    Adjusted R-squared: -0.6384

Orthogonal polynomials of degree higher than 12 do not give a solution.

3. **Polynomial regression between failures (TFN, Y) and annual Dst mean (ADSTM, X)**

*A. Raw Polynomials*

```
Coefficients:
                     Estimate Std. Error t value Pr(>|t|)
(Intercept)        -5.948e+01  6.538e+03  -0.009    0.993
```

poly(sample1$x, 10, raw = TRUE)1   3.212e+02  6.310e+03   0.051    0.960
poly(sample1$x, 10, raw = TRUE)2  -2.257e+02  2.486e+03  -0.091    0.929
poly(sample1$x, 10, raw = TRUE)3   7.073e+01  5.360e+02   0.132    0.897
poly(sample1$x, 10, raw = TRUE)4  -1.222e+01  7.089e+01  -0.172    0.866
poly(sample1$x, 10, raw = TRUE)5   1.277e+00  6.064e+00   0.211    0.837
poly(sample1$x, 10, raw = TRUE)6  -8.410e-02  3.422e-01  -0.246    0.810
poly(sample1$x, 10, raw = TRUE)7   3.507e-03  1.265e-02   0.277    0.787
poly(sample1$x, 10, raw = TRUE)8  -8.978e-05  2.949e-04  -0.304    0.766
poly(sample1$x, 10, raw = TRUE)9   1.287e-06  3.925e-06   0.328    0.749
poly(sample1$x, 10, raw = TRUE)10 -7.908e-09  2.274e-08  -0.348    0.735

Residual standard error: 13.14 on 11 degrees of freedom
Multiple R-squared: 0.3298,   Adjusted R-squared: -0.2795
F-statistic: 0.5413 on 10 and 11 DF,  p-value: 0.8285

Higher than 10 degree raw polynomials have singularities.

*B. Orthogonal Polynomials*

Coefficients:
              Estimate Std. Error t value Pr(>|t|)
(Intercept)      40.9091     3.5331  11.579  8.43e-05 ***
poly(sample1$x, 16)1  -11.3784    16.5716  -0.687    0.523
poly(sample1$x, 16)2    0.3165    16.5716   0.019    0.986
poly(sample1$x, 16)3    3.0056    16.5716   0.181    0.863
poly(sample1$x, 16)4    6.4684    16.5716   0.390    0.712
poly(sample1$x, 16)5  -10.5535    16.5716  -0.637    0.552
poly(sample1$x, 16)6   -7.2902    16.5716  -0.440    0.678
poly(sample1$x, 16)7    6.7988    16.5716   0.410    0.699
poly(sample1$x, 16)8   19.6704    16.5716   1.187    0.289
poly(sample1$x, 16)9  -11.6455    16.5716  -0.703    0.514
poly(sample1$x, 16)10  -4.5698    16.5716  -0.276    0.794
poly(sample1$x, 16)11  -5.7445    16.5716  -0.347    0.743
poly(sample1$x, 16)12  -6.1674    16.5716  -0.372    0.725
poly(sample1$x, 16)13  -6.3801    16.5716  -0.385    0.716
poly(sample1$x, 16)14  16.2125    16.5716   0.978    0.373
poly(sample1$x, 16)15  10.1829    16.5716   0.614    0.566
poly(sample1$x, 16)16  -6.9156    16.5716  -0.417    0.694
---
Signif. codes:  0 '***' 0.001 '**' 0.01 '*' 0.05 '.' 0.1 ' ' 1

Residual standard error: 16.57 on 5 degrees of freedom
Multiple R-squared: 0.5155,   Adjusted R-squared: -1.035
F-statistic: 0.3324 on 16 and 5 DF,  p-value: 0.9578

Orthogonal polynomials with degree larger than 16 do not give a solution.

## 4. Polynomial regression between failures (TFN, Y) and annual number of days with Ap ≥ 40 (ANDALF, X)

*A. Raw Polynomials*
Coefficients:

```
                                    Estimate   Std. Error  t value  Pr(>|t|)
(Intercept)                         4.470e+01  6.048e+00   7.391    1.38e-05
poly(sample4$xandalf, 10, raw = TRUE)1   8.385e+00  5.006e+01   0.167    0.870
poly(sample4$xandalf, 10, raw = TRUE)2  -1.329e+01  3.304e+01  -0.402    0.695
poly(sample4$xandalf, 10, raw = TRUE)3   5.746e+00  8.633e+00   0.666    0.519
poly(sample4$xandalf, 10, raw = TRUE)4  -1.056e+00  1.193e+00  -0.886    0.395
poly(sample4$xandalf, 10, raw = TRUE)5   1.021e-01  9.713e-02   1.052    0.316
poly(sample4$xandalf, 10, raw = TRUE)6  -5.719e-03  4.880e-03  -1.172    0.266
poly(sample4$xandalf, 10, raw = TRUE)7   1.919e-04  1.526e-04   1.258    0.235
poly(sample4$xandalf, 10, raw = TRUE)8  -3.802e-06  2.885e-06  -1.318    0.214
poly(sample4$xandalf, 10, raw = TRUE)9   4.096e-08  3.013e-08   1.359    0.201
poly(sample4$xandalf, 10, raw = TRUE)10 -1.847e-10  1.331e-10  -1.388    0.193
```

Multiple R-squared: 0.4321, Adjusted R-squared: -0.08423

Higher than 10 degree raw polynomials have singularities.

*B. Orthogonal Polynomials*

Coefficients:

```
                              Estimate   Std. Error  t value  Pr(>|t|)
(Intercept)                   40.90909   2.61236    15.660   2.76e-07 ***
poly(sample4$xandalf, 13)1   -13.46134  12.25307    -1.099    0.304
poly(sample4$xandalf, 13)2     8.05041  12.25307     0.657    0.530
poly(sample4$xandalf, 13)3    -0.09468  12.25307    -0.008    0.994
poly(sample4$xandalf, 13)4    -5.03311  12.25307    -0.411    0.692
poly(sample4$xandalf, 13)5     6.90350  12.25307     0.563    0.589
poly(sample4$xandalf, 13)6    -8.58225  12.25307    -0.700    0.504
poly(sample4$xandalf, 13)7    -9.78147  12.25307    -0.798    0.448
poly(sample4$xandalf, 13)8    19.59628  12.25307     1.599    0.148
poly(sample4$xandalf, 13)9    -8.38585  12.25307    -0.684    0.513
poly(sample4$xandalf, 13)10  -16.78443  12.25307    -1.370    0.208
poly(sample4$xandalf, 13)11  -11.63954  12.25307    -0.950    0.370
poly(sample4$xandalf, 13)12    4.69138  12.25307     0.383    0.712
poly(sample4$xandalf, 13)13  -15.83747  12.25307    -1.293    0.232
---
```
Multiple R-squared: 0.5762, Adjusted R-squared: -0.1126

Higher than 13 degree orthogonal polynomials do not give a solution.

## 5. Polynomial regression between failures (TFN, Y) and annual number of days with Dst $\leq -40$ (ANDSTLF, X)

*A. Raw polynomials*

Coefficients:

```
                               Estimate Std. Error t value Pr(>|t|)
(Intercept)                   5.469e+01  8.785e+00   6.225 9.82e-05 ***
poly(sample1$x, 11, raw = TRUE)1  -4.302e+00  1.348e+01  -0.319   0.7561
poly(sample1$x, 11, raw = TRUE)2  -3.177e+00  6.255e+00  -0.508   0.6226
poly(sample1$x, 11, raw = TRUE)3   1.057e+00  1.069e+00   0.989   0.3460
poly(sample1$x, 11, raw = TRUE)4  -1.205e-01  9.160e-02  -1.315   0.2178
poly(sample1$x, 11, raw = TRUE)5   7.030e-03  4.539e-03   1.549   0.1524
poly(sample1$x, 11, raw = TRUE)6  -2.400e-04  1.394e-04  -1.721   0.1159
poly(sample1$x, 11, raw = TRUE)7   5.064e-06  2.736e-06   1.851   0.0939 .
poly(sample1$x, 11, raw = TRUE)8  -6.685e-08  3.430e-08  -1.949   0.0798 .
poly(sample1$x, 11, raw = TRUE)9   5.370e-10  2.653e-10   2.025   0.0704 .
poly(sample1$x, 11, raw = TRUE)10 -2.397e-12  1.151e-12  -2.082   0.0639 .
poly(sample1$x, 11, raw = TRUE)11  4.552e-15  2.140e-15   2.127   0.0593 .
---
Signif. codes:  0 '***' 0.001 '**' 0.01 '*' 0.05 '.' 0.1 ' ' 1
```

Residual standard error: 9.141 on 10 degrees of freedom
Multiple R-squared: 0.7052, Adjusted R-squared: 0.3808
F-statistic: 2.174 on 11 and 10 DF, p-value: 0.1159

Raw polynomials with degree larger than 11 have singularities.

*B. Orthogonal Polynomials*

Coefficients:

```
                        Estimate Std. Error t value Pr(>|t|)
(Intercept)              40.909     1.576  25.950 2.16e-07 ***
poly(sample1$x, 15)1    -14.707     7.394  -1.989   0.0938 .
poly(sample1$x, 15)2      3.111     7.394   0.421   0.6886
poly(sample1$x, 15)3      9.037     7.394   1.222   0.2675
poly(sample1$x, 15)4     -1.849     7.394  -0.250   0.8109
poly(sample1$x, 15)5     -7.250     7.394  -0.981   0.3647
poly(sample1$x, 15)6     15.913     7.394   2.152   0.0749 .
poly(sample1$x, 15)7      5.364     7.394   0.725   0.4955
poly(sample1$x, 15)8      9.886     7.394   1.337   0.2297
poly(sample1$x, 15)9    -25.205     7.394  -3.409   0.0143 *
poly(sample1$x, 15)10    15.544     7.394   2.102   0.0802 .
poly(sample1$x, 15)11    19.443     7.394   2.630   0.0391 *
poly(sample1$x, 15)12    14.494     7.394   1.960   0.0977 .
poly(sample1$x, 15)13     4.764     7.394   0.644   0.5433
poly(sample1$x, 15)14     8.668     7.394   1.172   0.2855
poly(sample1$x, 15)15   -14.128     7.394  -1.911   0.1046
```

---
Signif. codes:  0 '***' 0.001 '**' 0.01 '*' 0.05 '.' 0.1 ' ' 1

Residual standard error: 7.394 on 6 degrees of freedom
Multiple R-squared: 0.8842,	Adjusted R-squared: 0.5948
F-statistic: 3.055 on 15 and 6 DF,  p-value: 0.08773

Orthogonal polynomials with degree larger than 15 do not give a solution.

6. **Polynomial regression between failures (TFN, Y) and annual mean of daily K-index sum (AMDKS, X)**

*A. Raw Polynomials*

Coefficients:

|  | Estimate | Std. Error | t value | Pr($>$\|t\|) |
|---|---|---|---|---|
| (Intercept) | 7.035e+05 | 8.463e+05 | 0.831 | 0.420 |
| poly(sample6$xamdks, 7, raw = TRUE)1 | -3.113e+05 | 3.813e+05 | -0.817 | 0.428 |
| poly(sample6$xamdks, 7, raw = TRUE)2 | 5.845e+04 | 7.289e+04 | 0.802 | 0.436 |
| poly(sample6$xamdks, 7, raw = TRUE)3 | -6.038e+03 | 7.670e+03 | -0.787 | 0.444 |
| poly(sample6$xamdks, 7, raw = TRUE)4 | 3.708e+02 | 4.798e+02 | 0.773 | 0.453 |
| poly(sample6$xamdks, 7, raw = TRUE)5 | -1.354e+01 | 1.785e+01 | -0.759 | 0.461 |
| poly(sample6$xamdks, 7, raw = TRUE)6 | 2.726e-01 | 3.659e-01 | 0.745 | 0.469 |
| poly(sample6$xamdks, 7, raw = TRUE)7 | -2.333e-03 | 3.188e-03 | -0.732 | 0.476 |

Multiple R-squared: 0.2597,	Adjusted R-squared: -0.1104

Raw polynomials with degree larger than 7 give singularities.

B. Orthogonal polynomials

Coefficients:

|  | Estimate | Std. Error | t value | Pr($>$\|t\|) |  |
|---|---|---|---|---|---|
| (Intercept) | 40.90909 | 3.29223 | 12.426 | 5.98e-05 | *** |
| poly(sample6$xamdks, 16)1 | -19.47957 | 15.44194 | -1.261 | 0.263 |  |
| poly(sample6$xamdks, 16)2 | -0.04309 | 15.44194 | -0.003 | 0.998 |  |
| poly(sample6$xamdks, 16)3 | 6.03978 | 15.44194 | 0.391 | 0.712 |  |
| poly(sample6$xamdks, 16)4 | 0.66714 | 15.44194 | 0.043 | 0.967 |  |
| poly(sample6$xamdks, 16)5 | -12.82891 | 15.44194 | -0.831 | 0.444 |  |
| poly(sample6$xamdks, 16)6 | 8.64415 | 15.44194 | 0.560 | 0.600 |  |
| poly(sample6$xamdks, 16)7 | -8.95980 | 15.44194 | -0.580 | 0.587 |  |
| poly(sample6$xamdks, 16)8 | 1.71347 | 15.44194 | 0.111 | 0.916 |  |
| poly(sample6$xamdks, 16)9 | 8.57691 | 15.44194 | 0.555 | 0.603 |  |
| poly(sample6$xamdks, 16)10 | -8.54220 | 15.44194 | -0.553 | 0.604 |  |
| poly(sample6$xamdks, 16)11 | -2.39759 | 15.44194 | -0.155 | 0.883 |  |

```
poly(sample6$xamdks, 16)12 -17.47020  15.44194 -1.131   0.309
poly(sample6$xamdks, 16)13  -1.47257  15.44194 -0.095   0.928
poly(sample6$xamdks, 16)14   9.95850  15.44194  0.645   0.547
poly(sample6$xamdks, 16)15 -18.14349  15.44194 -1.175   0.293
poly(sample6$xamdks, 16)16   3.82624  15.44194  0.248   0.814
---
Multiple R-squared: 0.5793,    Adjusted R-squared: -0.7671
```

Orthogonal polynomials with degree larger than 16 do not give a solution.

Our results of this "extreme polynomial regression" can be summarised in the following table:

Table 8: Higher degree polynomial regression results

| **Long term solar activity index** | **Polynomials** | | | |
|---|---|---|---|---|
| | (ordinary) Raw | | Orthogonal | |
| | Max allowable degree | $R^2$ | Max allowable degree | $R^2$ |
| *SSN* | 12 | 0.60 | 17 | 0.81 |
| *Annual Ap mean* | 10 | 0.22 | 12 | 0.30 |
| *Annual Dst mean* | 10 | 0.33 | 16 | 0.52 |
| *Annual no of days with Ap ≥ 40* | 10 | 0.43 | 13 | 0.58 |
| *Annual no of days with Dst ≤ - 40* | 11 | 0.71 | 15 | 0.88 |
| *Annual mean of daily K-index sum (Penteli)* | 7 | 0.26 | 16 | 0.58 |

The annual number of days with Dst $\leq$ - 40 long term index gives a rather impressive $R^2$ value of 0.88 using orthogonal polynomials (of 15$^{th}$ degree).